# Some properties of a long lifetime strongly-coupled molecular plasma produced by high Rydberg excitation of nitric oxide in a supersonic free jet

Jingwei Guo[1, 3], Mehrvash Varnasseri[1, 3], Mikko Riese[1, 2, 3] and Klaus Müller-Dethlefs[1, 3] *

[1)] The Photon Science Institute, [2)] Dalton Nuclear Institute, [3)] School of Chemistry, The University of Manchester, Alan Turing Building, Manchester M13 9PL, United Kingdom

A long life-time (>0.3 *ms*) strongly-coupled molecular Rydberg plasma is generated by the excitation of nitric oxide into the high-*n* Rydberg threshold region in the high-density region of a supersonic jet expansion. After 310 *µs* the plasma has expanded to a size of *ca.* 3 *cm*. When subjected to very small DC fields from 0.2 to 1.0 *V/cm* the plasma reveals a much smaller high-density core structure of only 0.6 *cm*. The molecular Rydberg plasma is observed over a broad range of excitation energies, from threshold down to Rydberg states as low as *n* = 19.

* Corresponding Author:  E-mail: K.Muller-Dethlefs@manchester.ac.uk

In a previous paper, we reported a novel mechanism for the generation of an ultra-cold molecular plasma with strongly-coupled properties by exciting the molecule *para*- difluorobenzene (*p*DFB)  into the high-n Rydberg region in the high-density, high-collision rate region of a pulsed supersonic jet at a distance z very close to the nozzle [1]. Under those conditions, *i.e* for an ion density of $10^{12}$ to $10^{14}$ $cm^{-3}$ and excitation of *ca.* 4 $cm^{-1}$ below the vacuum level, the electron Rydberg orbit radius of 1.4 *µm* (for a single Rydberg state, *n*≈165) is much bigger than the average distance of a few hundred *nm* between the cations. Since this ensemble of cations is produced collectively during excitation into the Rydberg region the electrons instantaneously loose the memory of their parent cation, resulting in the *direct* formation of a plasma. A plasma is characterized by the Coulomb coupling parameter $\Gamma_\alpha$ [2] where $\alpha=i$ stands for the ions and $\alpha=e$ for the electrons:

$\Gamma_\alpha = E_{Coulomb}/E_{thermal} = [e^2/(4\pi\varepsilon_0 a)]/k_B T_\alpha$
$= 2.69 \times 10^{-5} [n/10^{12} \ cm^{-3}]^{-1/3} [T_\alpha/10^6 \ K]^{-1}$   (1)

$E_{Coulomb}$ is the Coulomb interaction energy of the plasma, $E_{thermal}$ is the thermal energy of ions (or electrons), *e* is the electron charge, $a=(3/4\pi n)^{1/3}$ is the Wigner-Seitz radius and *n* is the plasma density.

The production of a strongly-coupled plasma by Rydberg excitation without prior cooling (e.g. laser cooling in a magneto optical trap, MOT) is possible under supersonic jet expansion conditions because i) the density of the ensemble of cations formed close to the valve's nozzle is much higher than the atom density in a MOT and ii) the collisional cooling of the ions in the supersonic jet expansion works against the disorder induced heating effect [3] and keeps the ions at the translational temperature of the supersonic jet of the order of tenth of *K*. Disorder induced heating in combination with three body recombination have so far prevented the observation of a plasma with $\Gamma_i$ significantly higher than unity for more than tens of microsecond in MOT experiments [4].

The plasma produced from *p*DFB molecules showed a hitherto unobserved ultra-long lifetime of ~0.5 *ms* and the Coulomb coupling parameter for the ions was estimated to be between 230 and 820. Although the temperature the electrons $T_e$ and their Coulomb coupling parameter $\Gamma_e$ were not explicitly determined this extreme long lifetime indicates that also the electrons were strongly coupled. The influence of small static and pulsed electric fields on the plasma showed interesting results, which at this point could not be explained without doubt. However, two possible explanations compatible with the observations were given: i) the plasma shows a shell structure and, ii) under the influence of a small electric field breaks-up into smaller partially charged parts, termed *"plaslets"*.

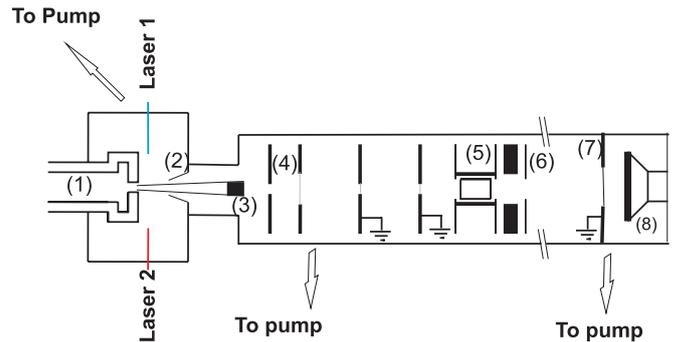

Figure 1: Schematic diagram of the experiment set-up. (1) General valve; (2) *6 mm* diameter skimmer; (3) plasma cloud; (4) extraction plates aperture 1 and grid 2; (5) X-, Y- deflectors; (6) ion lens; (7) rounded grid; (8) MCP detector

Completely independent from our work Grant and co-workers [5,6] have shown the possibility of producing an ultra-cold molecular plasma in a molecular beam (skimmed supersonic jet, 10.5 cm away from the nozzle) with observation times up to 30 $\mu s$. In their experiment NO was excited by a double resonant laser excitation process into high-n Rydberg states, which *indirectly* evolved into a plasma by avalanche Penning-type ionization. This is different to our plasma formation process where the plasma is formed *directly*.

In the present work we extend our investigations using a two-color two-photon (2C2P) process to study the plasma formation of NO in a Neon jet expansion, instead of a 1C2P process which was used for pDFB [1] (only possible due to coincidental resonances [7]); in contrast 2C2P process gives full control of intermediate state selection and Rydberg excitation energy. In addition, the experimental setup was changed from orthogonal acceleration to a collinear setup (see Figure 1) in order to allow a more detailed investigation of the central part of the plasma. A gas mixture of NO with Neon as carrier gas (1:5) is introduced into the vacuum chamber through a pulsed (General) valve with a *0.5 mm* nozzle at a backing pressure of 3 *bar* at a repetition rate of 20 *Hz*. The excitation of the molecules is performed by two frequency doubled Narrowscan dye lasers synchronously pumped by a Nd:YAG laser (Continuum 9020 Powerlite). Their outputs are overlapped in a counter-propagating way with the high-density region ($z=3.5$ mm downstream of the nozzle) of the supersonic jet expansion. Both dye lasers were calibrated with a wavelength meter (High Finesse WS Ultimate 487).

The molecular density in a distance z from the nozzle can be calculated by $n(z)=I(0)_{id}/(u(z) \times z^2)$, where $I(0)_{id}$ and $u(z)$ are the ideal centre-line intensity of a skimmerless source and the flow velocity, respectively [8,9]. For a pure Neon expansion the centreline intensity can easily be calculated, in analogy to the calculations for argon in Ref. 1, to $I(0)_{id} = 2.36\times10^{21}$ *molecules*$\cdot s^{-1} sr^{-1}$, resulting in a neon density of $2.25\times10^{17}$ $cm^{-3}$ and an NO molecule density of $4.5 \times10^{16}$ $cm^{-3}$ at a distance of $z=3.5$ *mm*.

The first dye laser (Coumarin 47), with an UV output of around *0.2 mJ* and a beam profile of *2 mm* in diameter, is used to excite the NO into the chosen rotationless intermediate state (A $^2\Sigma^+$ $v_A=0$, $N_A=0$, $J_A=1/2$) ← X $^2\Pi_{1/2}$ $v''=0$, $J''=1/2$) at 226.5 *nm* [10]. This intermediate state is used for all experiments presented here. The line intensities of a wavelengths scan of the first laser reflect a rotational temperature of $1K$-$2K$. At a temperature of 1.5 $K$ we have a fraction of 0.877 molecules in the $J''=1/2$ (±) states and a fraction of 0.44 in the $J''=1/2$ (-) state, corresponding to a density of $2\times10^{16}$ $cm^{-3}$ in this state.

The further excitation into the Rydberg region is performed with the second dye laser (DCM) of typical UV output of *ca. 5 mJ/pulse* (4 *mm* diam.) and the UV beam is slightly focused by a *f=200 mm* lens with the focus a few *cm* away from the jet. The intensity ratio of the two dye laser pulses was chosen so that only a two-color signal was observed.

Assuming an reasonable ionization efficiency of 10% we obtain a density of $NO^+$ ions of $2\times10^{15}$ $cm^{-3}$. Because in a typical jet expansion $T_{rot} \sim 10\times T_{trans}$ the translational temperature is estimated to around $0.15K$, which gives an expected ion Coulomb coupling parameter of around $2\times10^3$.

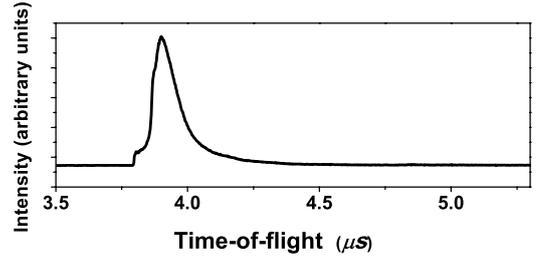

Figure 2: Observed time-of-flight spectrum of nitric oxide cation. A 3.6 *kV* pulsed voltage is applied on the first extraction plate 320 $\mu s$ after the laser is fired, the second extraction plate is grounded; following the excitation of the firstly, the second laser is fixed at 30516 $cm^{-1}$.

The supersonic jet has an average velocity of 0.85 $mm/\mu s$, so it takes the plasma *ca.* 310 $\mu s$ to fly to the first extraction plate (distance from the nozzle is 261 *mm*, hole *diam.* 4*mm*). Figure 2 shows the time-of-flight (TOF) signal observed when a 3.6 *kV* pulse is applied to the first extraction plate 320 $\mu s$ after the laser is fired (all other extraction plates are grounded). Here the energy of the second laser was set at 30,516 $cm^{-1}$, resulting in a two-photon energy of 4 $cm^{-1}$ below the adiabatic ionization energy [10]. The observed TOF distribution maps the spatial distribution of $NO^+$ ions between aperture 1 and 2 at the time when the ions are released from the plasma. It can be seen that the main signal originates from a position very close to aperture 1. A detailed study of this time structure will be reported later [11]. At present we are only presenting measurements for the total integrated signal as a function of the wavenumber of the second laser and of a small *DC* offset applied to the high voltage pulse on ap. 1.

Figure 3 shows the spectrum obtained by scanning the wavenumber of the second laser, while detecting the integrated ion signal. At lower excitation energies two Rydberg progressions converging to $X^+$ $^1\Sigma^+$, $N^+=0$ and $X^+$ $^1\Sigma^+$, $N^+=2$ are observed. In ZEKE experiments these states were observed [10] with an intensity ratio of 5:1. The smaller unassigned peaks may be originated from Rydberg states converging to states with different $N^+$ values. The distance between the first and second extraction plates is 3 *cm*, hence a pulse of 3.6 *kV* generates an electric field of 1200 *V/cm*. For single Rydberg states and diabatic ionization an electric field of the field strength *F* will decrease the ionization energy by $4\sqrt{F}$ [$cm^{-1}$] [12]. The electric field applied in this experiment is therefore only sufficient to ionize Rydberg states, which are not more than 140 $cm^{-1}$ below the ionisation threshold (corresponding to *n* = 23). However, in Figure 3 strong signals can clearly be observed down to 200 $cm^{-1}$ below the ionization threshold (*n* = 19). The lifetime of Rydberg states with n = 19, due to predissociation, is very much smaller than 320 *μs*. Therefore the initially produced states have to undergo changes to survive the flight time needed to reach aperture 1. It is expected that the initially prepared Rydberg states undergo resonant Rydberg collisions, which will increase the quantum number of one collision partner while reducing the quantum number of the other partner [13,14]. Multiple Rydberg collisions will generate long-lifetime high-n Rydberg states and also ions, which then form the plasma.

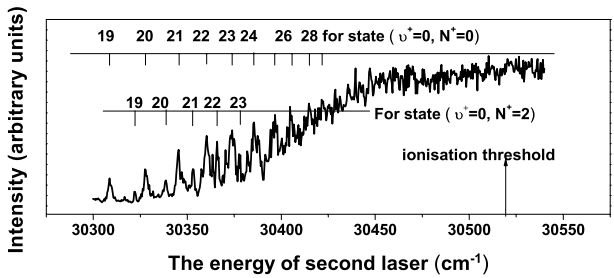

Figure 3: Excitation spectrum obtained by scanning the wavenumber of the second laser, while detecting the integrated $NO^+$ signal.

In an additional experiment the composition of the ion signal was investigated in a more detailed way. The wavelength of the second laser was fixed to 30516 $cm^{-1}$ and the time delay between the high voltage pulse and the laser pulse was scanned while the integrated ion signal was recorded, as displayed in Figure 4 a. A small DC offset from 0 to 5 *V* was applied to the *HV* pulse. Under the Ne supersonic jet condition, nitric oxide has a typical kinetic energy of 100 *meV*, therefore a DC offset of 1 *V* is more than sufficient to stop all single cations from passing through aperture 1. The small offset has a different effect on the plasma, which is overall neutral and will therefore not be stopped by this field. Under a weak electric field, the plasma will first loose the most outer loosely bound electrons followed by the ions. After the outermost ions are removed from the plasma, the inner ions are exposed to the electric field and also start to be removed. This process will be repeated until the whole plasma has eventually decayed

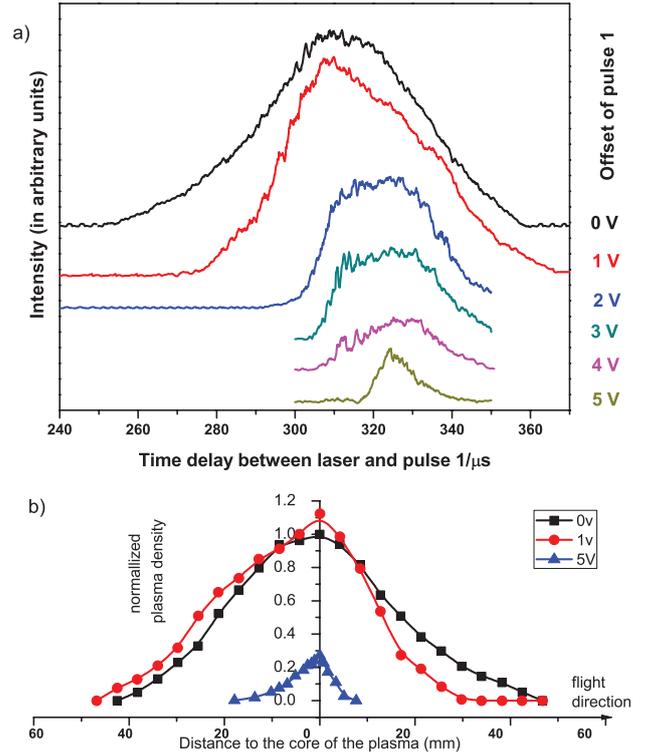

Figure 4: Time delay scans ($t_{pulse\ 1}$ - $t_{laser}$) spectra for different *DC* offset voltages of *0, 1, 2, 3, 4* and *5V* (corresponding to a maximum field strength of 0.0, 0.2, 0.4, 0.6, 0.8 and 1.0 *V/cm*) applied to the baseline of the high voltage pulse applied to aperture 1

into single cations. The denser the plasma, the longer it can survive in the same electric field. Without *DC* voltage the plasma signal can be observed for a delay between laser pulse and the high voltage pulse from ca. 250 *μs* to 360 *μs*. If an offset of 1 *V* is applied the signal starts to appear at later delay times (around 280 *μs*). While this field has a strong influence on the early part of the plasma the later part is hardly affected. The plasma can still be observed up to a delay time of 360 *μs*. Also the intensity of the peak maximum is not reduced. Increasing the offset by another volt has a drastic effect on the plasma. Now the early as well as the later part are affected in a similar way. The signal can now only be observed from ca. 300 to 340 *μs* and the intensity is reduced. Applying a voltage of 5 *V* reduces the size further, while the intensity is hardly decreased. The plasma can now only be detected over a time interval of 20 *μs*. Without offset the peak maximum can be found at around 310 *μs* while it is shifted to 325 *μs* when a offset of 5 *V* is applied. One explanation is that the plasma is slightly positively charged, and moves like a plasma crystal with a defined centre of mass. An increase of the offset voltage slows the plasma crystal only slightly down due to the extremely big mass to charge ratio, resulting in a slightly longer time-of-flight to arrive at aperture 1.

It is also possible that, when electrons are removed from the early part of plasma by the DC electric field, the free ions will fly towards the main body of the plasma. These ions either fly into the main body of plasma or are being diverted to other directions, giving the main body of the plasma a small momentum kick resulting in slowing it down. When the ions are pushed into the plasma, this can lead to a partially charged plasma, which is slowed down by the DC field as well. This is just a first idea and not yet a proven theory.

Figure 4b shows the spatial density distribution of the plasma for an offset of 0 *V*, 1 *V* and 5 *V* as extracted from figure 4a. The density is normalized with respect to the maximum signal without offset. Without offset the total plasma has a size (FWHM) of *ca.* 3 *cm*. For 1 *V* it can be seen that the density decreases in the first part of the plasma while the density of the maximum and also the later part is slightly increased. This is in accordance with the explanation that this small *DC* Voltage only removes the electrons of the first part of the plasma. The resulting ions are then accelerated against their flight direction. At 5 *V* the size of the plasma and the intensity is significantly reduced. The size is now only .6 *cm,* which can be interpreted as a dense plasma core. At this field strength electrons are removed from all over the outer part of the plasma. But this effect is still more effective for the first part, as the later part is shielded by the plasma. Therefore the density in the first part will decrease more than in the later part, which can be clearly seen from figure 4b.

It is useful to compare our results to the results obtained from Grant and co-workers [5,6]. They have observed that the application of a relatively short (1 μs) electrostatic pulse up to 150 V/cm shortly after plasma formation does hardly affect the signal associated with the plasma [5]. We observed that already very weak static fields (0.2 *to 1 V/cm*) strongly reduce the size of the plasma. There are two differences in the performed experiments. First, in our case the electric field is applied at much later times and, second, for a much longer time. It would therefore be an interesting point to investigate the effect of a short pulse with higher amplitude compared to a static field with low amplitude. We expect that the length of time the plasma is exposed to the field is more crucial than the field strength itself, but this will need to be part of further investigations. It is also possible that the small field can act on the plasma because our plasma had much more time to expand.

Further we want to compare the expansion of our plasma to that produced by Grant and co-workers [6]. The expansion of a plasma is driven by the electron temperature, which plays a crucial role [15]. The expansion of a quasi-neutral ultra cold plasma with a spherically symmetric Gaussian distribution function was described by Laha and coworkers [16] and this ansatz was chosen by Grant and co-workers to describe their expansion process. The characteristic expansion time $\tau_{exp}$ is given by:

$$\tau_{exp} = \sigma(0)[m_i/k_B(T_i(0)+T_e(0))]^{1/2} \quad (2)$$

where $T_\alpha(0)$ are the initial ion and electron temperatures, $m_i$ is the ion mass and $\sigma(0)$ is the initial plasma diameter. As the diameter expands with the time according to:

$$\sigma(t) = \sigma(0)[1+t^2/\tau_{exp}^2]^{1/2} \quad (3)$$

the temperature of the ions and electrons falls as given by: $T_\alpha(t) = T_\alpha(0)/[1+t^2/\tau_{exp}^2]$. (4)

The plasma from Grant and co-workers expanded from an initial width of 785 *μm* to 1617 *μm* in 28.7 *μs*. Using equations 2-4, an initial width of 2 *mm* and an expansion time of 310 *μs*, an expansion to a size of 4.4 *cm* would result, which is significantly more than observed here. We had expected an even slower expansion because the plasma formed in our experiment is expected to be much more strongly coupled. In a forthcoming paper we report the interesting result that our plasma does not behave like a gas and gets compressed on the first aperture [11]. This observation also has implications on how to determine the real plasma size as a function of expansion time.

M. R. thanks The University of Manchester-Nuclear Decommissioning Authority Dalton Cumbria Project for support of this work.